# Anisotropic Thermal Conductivity in Single Crystal β-Gallium Oxide

Zhi Guo,[1,6] Amit Verma,[2] Fangyuan Sun,[3] Austin Hickman,[2] Takekazu Masui,[4] Akito Kuramata,[4] Masataka Higashiwaki,[5] Debdeep Jena[2] and Tengfei Luo[1,*]

[1]*Department of Aerospace and Mechanical Engineering, University of Notre Dame, Notre Dame, IN 46556, USA*

[2]*Department of Electrical Engineering, University of Notre Dame, Notre Dame, IN 46556, USA*

[3]*Institute of Engineering Thermophysics, Chinese Academy of Sciences, Beijing, 100190, China*

[4]*Tamura Co., Ltd., 2–3–1 Hirosedai, Sayama, Saitama 350–1328, Japan*

[5]*National Institute of Information and Communications Technology, 4–2–1 Nukui-kitamachi, Koganei, Tokyo, 184-8795, Japan*

[6]*Radiation Laboratory, University of Notre Dame, Notre Dame, IN 46556, USA*

The thermal conductivities of β-Ga$_2$O$_3$ single crystals along four different crystal directions were measured in the temperature range of 80-495K using the time domain thermoreflectance (TDTR) method. A large anisotropy was found. At room temperature, the [010] direction has the highest thermal conductivity of 27.0±2.0 W/mK, while that along the [100] direction has the lowest value of 10.9±1.0 W/mK. At high temperatures, the thermal conductivity follows a ~1/T relationship characteristic of Umklapp phonon scattering, indicating phonon-dominated heat transport in the β-Ga$_2$O$_3$ crystal. The measured experimental thermal conductivity is supported by first-principles calculations which suggest that the anisotropy in thermal conductivity is due to the differences of the speed of sound along different crystal directions.

**I. INTRODUCTION**

β-Ga$_2$O$_3$ is a wide-bandgap semiconductor with a bandgap of ~4.8eV[1]. Because of the large bandgap and the resultant large electrical breakdown strength, this material can sustain large voltages, making it attractive for high voltage device applications. Recently, β-Ga$_2$O$_3$ Schottky diodes with large reverse breakdown voltages and β-Ga$_2$O$_3$ field-effect transistors sustaining large drain voltages have been demonstrated[1-4]. In a high-voltage device, most of the power is dissipated in the channel, causing an increase in the channel temperature. Such Joule heating can increase the channel temperature by tens or even over one hundred degrees above ambient temperature if heat is not removed efficiently. High temperatures result in the degradation of the electron transport properties due to increased electron-phonon scattering, resulting in the loss of device performance. Efficient removal of the generated heat is required for maintaining the performance, and also for ensuring the reliability of the device. β-Ga$_2$O$_3$ crystallizes in a highly anisotropic monoclinic crystal structure with the space group C2/m and lattice constants of *a* = 12.214 Å, *b* = 3.0371 Å, *c* = 5.7981 Å, and *β*=103.83 °[5]. Because of this crystalline anisotropy, the thermal conductivity in β-Ga$_2$O$_3$ is expected to be different along different crystal directions. The knowledge of the anisotropic thermal transport in β-Ga$_2$O$_3$ at intermediate and high temperatures is of high importance for high-voltage device design for optimal thermal management. However, currently this anisotropic aspect of thermal transport in β-Ga$_2$O$_3$ is poorly understood.

Handwerg et al. have recently reported the temperature-dependent thermal conductivity of Mg-doped and undoped β-Ga$_2$O$_3$ bulk crystals from 4K to room temperature using 3ω-method[6]. The thermal conductivity determined was along the [100] direction. Along this direction, the thermal conductivity at room temperature was found to be 13.0±1.0 W/mK and a peak thermal conductivity of about $(5.3\pm0.6)\times10^2$ W/mK was found at 25 K. Laser-flash methods were used previously to measure thermal conductivity values of 13 W/mK along the [100] direction[7] and 21 W/mK along the [010] direction[8]. These values are for room temperature. Typical high-voltage device design requires knowledge of the high-temperature thermal conductivity, which is not yet explored.

In this work, we use time domain thermoreflectance (TDTR) to study the thermal conductivity of β-Ga$_2$O$_3$ single crystals along four different crystal orientations: [001], [100], [010], and [-201] over a large temperature range, including much higher temperatures than previous studies. We find a significant anisotropy along the four directions at all temperatures. The [100] direction has the smallest thermal conductivity among all directions, at 10.9±1.0 W/mK at room temperature. The thermal conductivity along the [010] direction is significantly higher - the value 27.0±2.0 W/mK is about three times of that along the [100] direction. The temperature-dependent thermal conductivities at high temperature for all directions show a 1/T relationship, indicating the phonon-dominant nature of the thermal transport.

**A. Sample growth and fabrication**

For this study, bulk β-Ga$_2$O$_3$ crystals with 4 crystal orientations ([001], [100], [010], and [-201]) were obtained from Tamura Corporation (Japan). These crystals were grown by an edge-defined film fed growth (EFG) method. The bulk single-crystals were doped with Sn during growth, resulting in a room-temperature carrier concentration of ~5-6 x 10$^{18}$ cm$^{-3}$. The crystal orientations of the samples were verified using X-ray diffraction measurements (see supporting information). The X-ray diffraction measurements directly verify the high crystallinity and orientation of the β-Ga$_2$O$_3$ single crystals. A 100 nm-thick Al film (nominal thickness) was then deposited using electron beam evaporation on top of the Ga$_2$O$_3$ crystals for TDTR measurements. The actual film thickness of was 103±5 nm determined by profilometry (KLA-Tencor, P6).

**B. TDTR measurement method and heat capacity measurement**

The TDTR system used for the thermal conductivity measurements was described elsewhere[9, 10]. The 1/e$^2$ radii (where beam diameter is defined by the intensity falls to 1/e$^2$ = 0.135 times the maximum value) of the pump and probe beams incident on the Al film surface were 30 μm and 5 μm, respectively. The modulation frequency of the pump beam was set to 5

MHz for all measurements. For temperature-dependent measurements, a liquid nitrogen-cooled cryogenic system (ST-100, Janis Research) with a temperature-control module (Lakeshore Ltd.) was used to cool or to heat the sample and stabilize the sample at target temperatures. To monitor the actual temperature on the sample surface, a calibrated silicon diode (DT-670-SD) was attached to the sample surface and connected to the temperature controller. The setup described above allows us to investigate the temperature range from 80 K to 495 K with a 0.01 K precision.

To extract the thermal conductivity from the thermoreflectance decay data, the phase signal data demodulated from the lock-in amplifier was used to fit a pulse-accumulated heat conduction model[11]. To obtain the thermal conductivity, it is necessary to know the heat capacity of $Ga_2O_3$ at different temperatures. Since this data is not available in the literature, we have performed the measurement of heat capacity of β-$Ga_2O_3$ crystals in the temperature range of 123 K to 748 K, using a differential scanning calorimeter (Mettler Toledo, DSC-1). The measured temperature-dependent specific heat capacity is shown in Figure 1. The data are fitted to the Debye model of the specific heat

$$C(T) = 3Nk \frac{3}{x_0^3} \int_0^{x_0} \frac{x^4 e^x}{(e^x - 1)^2} dx \qquad (1)$$

where $x_0 = \frac{\hbar \omega_D}{kT} = \frac{\theta_D}{T}$. Here $q_D$ is the Debye temperature, $T$ is the absolute temperature, $\hbar$ is the reduced Planck's constant, and $k$ is the Boltzmann constant. In Eqn.1, $N$ is the number of atoms in the sample. Since, each formula unit of $Ga_2O_3$ has 5 atoms, each mole of $Ga_2O_3$ formula units (~187.438 g of β-$Ga_2O_3$) has 5 moles of atoms. Measured data as shown in Figure 1 has been normalized to one mole of atoms (N=6.023 x $10^{23}$). From the fit in Figure 1, the Debye temperature is obtained as 738 K, in reasonable agreement with a predicted value of 872 K from first-principles calculations[12]. With these values of the heat capacity, the thermal conductivity is extracted from the TDTR measurements.

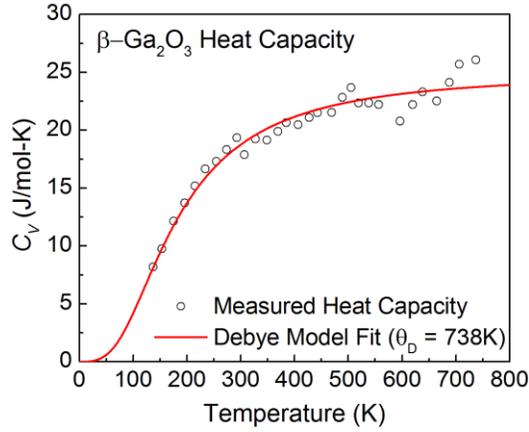

FIG. 1. Measured heat capacity of β-$Ga_2O_3$ as a function of temperature. The solid line is the Debye model fit from which the Debye temperature is obtained.

## II. Results and discussion

Since the laser beam sizes used for our TDTR measurements are much larger than the thermal penetration depth, one-dimensional heat transfer into the surface can be assumed. As a result, samples with different crystalline orientations allow us to directly probe the thermal transport along specific crystal directions. Through TDTR measurements, the temperature-dependent thermal conductivity measured along different directions at various temperatures are shown in Figure 2. The thermal conductivity is the highest along the [010] direction, and lowest along the [100] direction at all temperatures used in the measurements. The thermal conductivity in the [100] direction is consistent with Handwerg's report[6], with the value increasing from ~10 W/mK by roughly an order of magnitude around 100K. The thermal conductivity along crystal orientations of smaller lattice constant is larger, as indicated in the inset of Figure 2a.

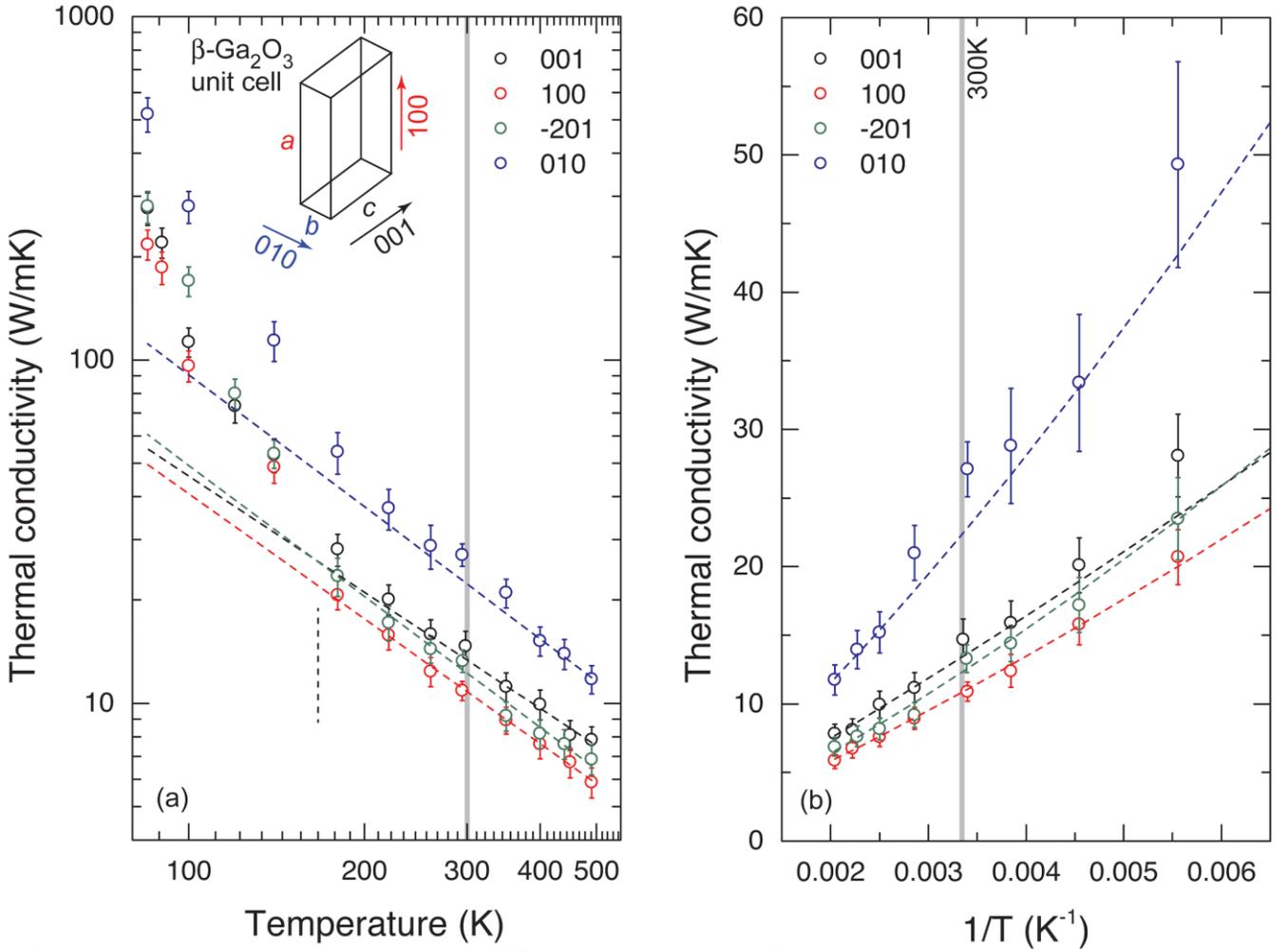

FIG. 2. Temperature-dependent thermal conductivity of β-Ga$_2$O$_3$ measured along different crystal directions by the TDTR approach. In (a), the thermal conductivity and temperature are in the log scale. The inset shows a schematic of the unit cell of the β-Ga$_2$O$_3$ crystal. The thermal conductivity is larger along directions of smaller lattice constant: the rough lattice constant ratios are c~2b and a~4b. The dashed lines show $1/T^m$ fits that capture the high-temperature behavior of the thermal conductivity. The vertical dashed line separates the high-temperature behavior from the lower-temperature deviation to the fits. (b) Shows a linear plot of thermal conductivity against $1/T$ to highlight the dependence on temperature and the high-temperature $1/T^m$ fits more clearly.

As a wide-bandgap semiconductor, a major part of the thermal conductivity of Ga$_2$O$_3$ is expected to be due to crystal vibrations (phonons) with a small contribution from free carriers. Although the samples are doped with Sn with a concentration of ~5-6 x $10^{18}$ cm$^{-3}$, the mobile electron contribution to thermal conductivity is negligible when the carrier concentration is lower than $10^{19}$ cm$^{-3}$ [13]. At high-temperatures approaching the Debye temperature, umklapp scattering is the dominant phonon scattering mechanism in a near-perfect crystal with low density of defects. Therefore, we focus our analysis in the high temperature region. In this regime, the specific heat capacity $c_v = \frac{C_v}{V}$ of phonons saturates according to the Dulong-Petit law: Figure 1 shows the heat capacity approaching a high-temperature saturation value of ~25 J/mol-K. The thermal conductivity $\kappa \simeq \frac{1}{3}(v_s l)c_v$ at high temperatures depends on the sound velocity $v_s$ and the mean-free path $l$ of phonons. Because the relaxation time of acoustic phonons decreases as 1/T due to the increase in the phonon density, according to kinetic theory, the thermal conductivity at high temperatures should follow a 1/T relationship. In practice, this relationship often exhibits $1/T^m$ (m=1~1.5) behavior[14]. We fit the high temperature data (>350 K) to a $1/T^m$ relation in Figure 2 as dashed lines. As can be seen, the relation fits the thermal conductivities reasonably well at high temperatures. The fit extends down to 200K, well below the Debye temperature. For convenience in device modeling with a few parameters, we fit the thermal conductivity in the whole temperature range using $\kappa(T) = AT^{-m}$ in two different temperature regimes (80-200 K and 200-495 K), and present the parameters in Table I. As shown in Table I, for high temperature region (200-495 K), the index *m* in $1/T^m$ are all around 1 for all the directions.

Table I. A functional form of $\kappa(T) = AT^{-m}$ is used to fit the temperature dependent thermal conductivity data in two temperature ranges (range 1: 80-200 K; range 2: 200-495 K). The fitting value of each parameter in the equation has been summarized in fours columns (subscripts represent different temperature region).

| Crystallographic Orientation | $A_1$ | $m_1$ | $A_2$ | $m_2$ |
|---|---|---|---|---|
| [001] | $1.06 \times 10^{10}$ | 3.93 | $8.14 \times 10^3$ | 1.12 |
| [100] | $1.85 \times 10^9$ | 3.59 | $1.06 \times 10^4$ | 1.21 |
| [010] | $7.99 \times 10^8$ | 3.21 | $3.28 \times 10^4$ | 1.27 |
| [-201] | $8.26 \times 10^8$ | 3.35 | $1.69 \times 10^4$ | 1.28 |

If we assume a gray model for phonon transport (i.e., all phonons have the same group velocity and relaxation time), the thermal conductivity in a specific direction, *i*, can be expressed as[15,14]

$$\kappa_i = c_v (v_i \cdot v_i) \tau \qquad (2)$$

where $c_v$ denotes the volumetric heat capacity, $v_i$ is the group velocity in the direction *i* (=x,y,z), and $\tau$ is relaxation time. In the gray model, the phonon group velocity is usually approximated by the sound speed. Since both $c_v$ and $\tau$ are scalars, the orientation-dependent group velocity $v$ thus solely determines the anisotropy of the thermal conductivity. This is also confirmed by experiments: the thermal conductivity along different directions can be approximately related to each another by a multiplication factor at high temperature (>200 K). Such a factor is related to the magnitude of the square of the sound speed in different directions. We can therefore obtain the relative sound speeds (normalized to that in the [001] direction) from the temperature dependent thermal conductivity data (Table II).

We further calculate the phonon dispersion relation of β-Ga$_2$O$_3$ using Density Functional Perturbation Theory (DFPT)[16] to extract the phonon group velocities at different directions to justify the experimental findings. In Figure 3, we plot the acoustic phonon dispersion relationship calculated from a 10-atom primitive cell along four specified directions using the PWscf and PHonon modules of the Quantum Expresso package[17]. The Perdew-Burke-Ernzerhof functional, and norm-conserving pseudopotentials were used. The plane-wave basis kinetic energy cut-off was set to 60 Ry. A 5×5×3 k-point mesh was used for Brillouin zone sampling. We note that the phonon band structure obtained in this work are in good agreement with previous first-principles calculations on β-$Ga_2O_3$[18], but the dispersion relationship along the specific directions required here was not presented in the prior work. To approximate the sound speed along each specific direction of interest, we calculate the slopes of each acoustic branch in that direction near the Brillouin zone center and take the average of the square, i.e. $\bar{v}^2 = \frac{1}{3}(v_L^2 + v_{T1}^2 + v_{T2}^2)$.

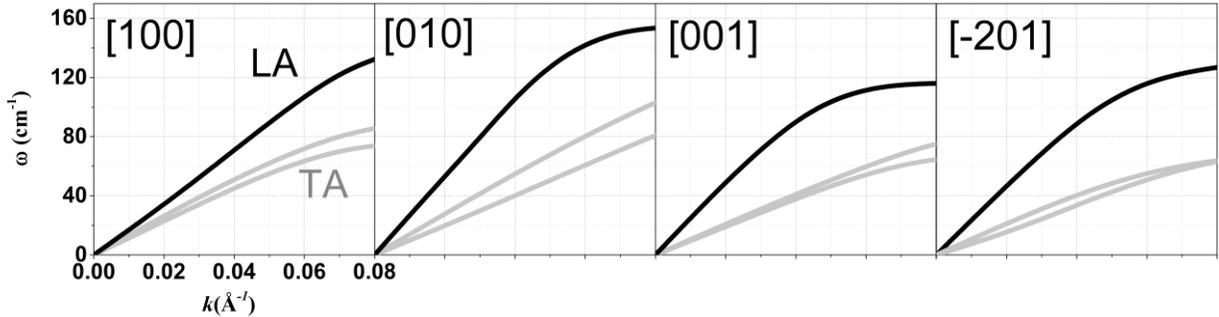

FIG. 3. Phonon dispersion relationship calculated along four different directions. For clarity, in each direction, only three acoustic phonon branches are shown. The LA modes are shown in black, while the TA modes are colored in grey. The Miller indices use the unit cell lattice vector definitions in order to compare with the experimental findings.

TABLE II. A comparison of the relative sound speed determined from thermal conductivity measurements and predicted sound speed from first-principles calculations. The sound speed values are normalized to the value along [001] direction. The linear regime of figure 3 was fitted and used to obtain the group velocity of each branch. $V_L$ represents the LA phonon group velocity and $V_T$ denotes the group velocity of two TA phonon branches.

| Crystallographic Orientation | Relative sound speed determined by experiments | Approximate mean sound speed from first-principles calculations | Absolute group velocity from first-principles calculations ($10^5$ cm/s) |
|---|---|---|---|
| [001] | 1.00 | 1.00 | 7.1 ($V_L$) 2.8~3.1 ($V_T$) |
| [100] | 0.88±0.2 | 0.91 | 5.4 ($V_L$) 3.4~3.9 ($V_T$) |
| [010] | 1.32±0.25 | 1.13 | 7.8 ($V_L$) 3.0~4.1 ($V_T$) |
| [-201] | 0.94±0.26 | 0.93 | 6.6 ($V_L$) 2.4~3.1 ($V_T$) |

In Table II, we find that the predicted phonon speeds from first-principles calculations show a very similar trend as that extracted from the experimental data. However, there are quantative differences. This can be attributed to two reasons. (1) Strictly speaking, the relative speed extracted using Eqn. 2 from the experimental data is not the speed of sound. Instead, what we obtain is the effective phonon group velocity weighted by factors related to the specific heat and relaxation time of all the phonon modes. (2) We have ignored the contributions of optical phonons to the thermal conductivity when using the acoustic phonon group velocities the first-principles calculations to approximate sound speeds. Nevertheless, since the acoustic phonons with long wavelengths are usually major contributors to the overall thermal conductivity in semiconductor crystals[19-23], we are still able to qualitatively explain the experimentally observed trend using the first-principles phonon group velocities.

It is interesting to compare the thermal conductivity of $Ga_2O_3$ to GaN, which has been a widely used semiconductor for power electronic applications. GaN has a thermal conductivity of 130-230 W/mK at room temperature,[24-26] and isotopically enriched GaN can have thermal conductivity up to 400 W/mK according first-principles calculations[27]. These values are much higher than that of the $Ga_2O_3$ which is in the range of 10-30 W/mK. According to the gray model (Eqn. 2), difference in $c_v$, $v$ and $\tau$ can lead to the difference in thermal conductivity between $Ga_2O_3$ and GaN. However, we found that the LA and TA mode group velocities of β-Ga2O3 (Table II) are not significantly different from the reported values of GaN ($V_L=6.9\times10^5$~$8.2\times10^5$ cm/s and $V_T=3.3\times10^5$~$5.0\times10^5$ cm/s along different directions)[28], and meanwhile their volumetric heat capacities differ by less than 20%[29]. It implies that a phonon relaxation time $\tau$ in GaN is much larger than that in $Ga_2O_3$. This conclusion is supported by a previous first-principles thermal conductivity calculations[27, 30]: The high thermal conductivity of GaN is attributed to the weak phonon-phonon umklapp scattering due to a large gap in its phonon band struture compared to other gallium compounds. In $Ga_2O_3$, such a gap is significantly smaller, and thus the phonon scattering is expected to be stronger, which lead to a much shorter relaxation time.

In conclusion, we find highly anisotropic thermal conductivity in β-$Ga_2O_3$ single crystals using TDTR measurements over a large temperature range, specifically at high temperatures. Among the directions investigated, the [010] direction has the highest thermal conductivity and that along the [100] direction is the lowest. At high temperatures, the thermal conductivity generally follows a 1/T relationship, indicating a phonon-dominated thermal transport. The anisotropy in the phonon group velocities explains the observed thermal conductivity anisotropy. The thermal conductivity data will be helpful for the thermal management and design of $Ga_2O_3$-based devices.

**ACKNOWLEDGMENTS**

We thank the Center for Sustainable Energy at Notre Dame (cSEND) Materials Characterization Facilities for the use of the DSC. This work was supported in part by the Sustainable Energy Initiative (SEI) at the University of Notre Dame and the Semiconductor Research Corporation (contract number: 2013-MA-2383). The computation in this research was supported in part by the Notre Dame Center for Research Computing and NSF through TeraGrid resources provided by TACC Stampede under grant number TG-CTS100078.